\documentclass[submission,copyright,creativecommons,english]{eptcs}
\providecommand{\event}{DCM 2011} 
\usepackage{breakurl}             
\usepackage[T1]{fontenc}
\usepackage[utf8]{inputenc}
\usepackage{amsthm}
\usepackage{amsmath}
\usepackage{amssymb}
\usepackage{latexsym}
\usepackage{stmaryrd}
\usepackage{algorithm}
\usepackage{multirow}
\usepackage{amstext}
\usepackage{graphicx}
\usepackage{amssymb}
\theoremstyle{plain}
  \newtheorem{thm}{Theorem}
  
  \newtheorem{prop}[thm]{Proposition}
\theoremstyle{definition}
   \newtheorem{defn}[thm]{Definition}

\usepackage{versions}\excludeversion{ignore}\excludeversion{comment}
\usepackage[usenames]{color}

\newcommand{\Gr}[1]{{\widetilde #1}} 
\newcommand{\F}{{F}}

\newcommand{\M}{\mathcal{M}}
\newcommand{\In}{{I}}
\newcommand{\If}{\mbox{\bf if}\;}
\newcommand{\Then}{\;\mbox{\bf then}\;}
\newcommand{\Else}{\;\mbox{\bf else}\;}
\newcommand{\True}{\textsc{true}}

\newcommand{\Undef}{\textsc{undef}}

\title{A Formalization and Proof of the \\Extended Church-Turing Thesis\\[1ex]
\Large---Extended Abstract---}
\def\titlerunning{Extended Church-Turing Thesis}
\def\authorrunning{N. Dershowitz \&  E. Falkovich}
\author{Nachum Dershowitz 
\institute{School of Computer Science\\Tel Aviv University\\Tel Aviv, Israel}
\email{nachum.dershowitz@cs.tau.ac.il}
\and Evgenia Falkovich%
\footnote{This work was carried out in partial fulfillment of the requirements for the Ph.D.\ degree
of the second author.}
\institute{School of Computer Science\\Tel Aviv University\\Tel Aviv, Israel}
\email{jenny.falkovich@gmail.com}}
\date{\today}
\usepackage{babel}
\begin{document}
\maketitle
\begin{abstract}
We prove the \textit{Extended Church-Turing Thesis}: Every effective
algorithm can be efficiently simulated by a Turing machine. 
This is accomplished by emulating an effective algorithm via an abstract state machine,
and simulating such an abstract state machine by a random access machine,
representing data as a minimal term graph.
\end{abstract}

\section{Introduction}

The Church-Turing Thesis
asserts that all effectively computable numeric functions are recursive and, likewise,
they can be computed by a Turing machine,
or---more precisely---can be simulated under some representation by a Turing machine.
This claim has recently been axiomatized and proven~\cite{CT_ASM,BSL}.
The ``extended'' thesis adds the belief that 
the overhead in such a simulation is polynomial.
One formulation of this extended thesis is as follows:

\begin{quote}
The Extended Church-Turing Thesis states \dots\  that time on all
{}``reasonable'' machine models is related by a polynomial. (Ian Parberry~\cite{Parberry})
\end{quote}

We demonstrate the validity of this thesis for all (sequential, deterministic, non-interactive) effective models over arbitrary constructive domains in the following manner:
\begin{enumerate}
\item We adopt the axiomatic characterization of (sequential) \textit{algorithms}
over arbitrary domains due to Gurevich~\cite{ASM-Theorem-Gurevich} (Section~\ref{sec:alg}, Definition~\ref{def:alg}). 
\item We adopt the formalization of \textit{effective} algorithms over arbitrary domains
from~\cite{CT_ASM} (Section~\ref{sub:Eff}, Definition~\ref{def:eff}). 
\item We adopt the definition of \textit{simulation} of algorithms in different models of computation given in~\cite{Simulation}.
\item We consider \textit{implementations}, which are algorithms operating
over a specific domain (Section~\ref{sec:alg}, Definition~\ref{def:imp}). 
\item We represent domain elements by their minimal constructor-based graph (dag) representation;
cf.~\cite{dagRepr} (Section~\ref{sec:com}). 
\item We measure the size of input as the number of vertices in a constructor-based
representation (Section~\ref{sec:com}, Definition~\ref{def:size}). 
\item We emulate effective algorithms step-by-step by abstract state machines~\cite{ASM-Theorem-Gurevich}
in the precise manner of~\cite{Exact} (Section~\ref{sec:asm}, Section~\ref{def:asm}).
\item Each basic implementation step can be simulated in a
linear number of random-access machine (RAM) steps (Section~\ref{sec:sim}, Theorem~\ref{Simulating-the-transition}). 
\item Input states to the simulation can be encoded
in linearly many RAM steps (Section~\ref{sec:sim}, Theorem~\ref{Simulating-initial-state}). 
\item As multitape Turing machines simulate RAMs in quadratic time \cite{RAM},
the thesis follows (Section~\ref{sec:sum}).
\end{enumerate}

\section{Algorithms}\label{sec:alg}\label{sub:Eff}

\begin{ignore}

We are interested in comparing the time complexity of algorithms implemented in different effective models of computation, models that may take advantage of
arbitrary data structures.
The belief that all feasible models make relatively similar demands on resources
when executing algorithms underlies the study of complexity theory.
\end{ignore}

First of all, an algorithm, in its classic sense, is a time-sequential
state-transition system, whose transitions are partial functions
on its states.
This ensures that each state is self-contained and that the next state, if any,
is determined.
The necessary information in states can be captured using logical structures, and
an algorithm is expected to be independent of the choice of representation
and to produce no unexpected elements. 
Furthermore, an algorithm should possess a finite description. 

\begin{defn}[Algorithm \cite{ASM-Theorem-Gurevich}]\label{def:alg}
A \emph{classical algorithm}
is 
a (deterministic) state-transition system, 
satisfying the following three postulates:
\begin{enumerate}
\item[{I}.] \label{I}
It is comprised of a set%
\footnote{Or class---it doesn't matter.}
$S$ of \emph{states}, a subset $S_{0}\subseteq S$ of \emph{initial}
states, and a partial \emph{transition} function $\tau:S\rightharpoonup S$
from states to states. 
States for which there is no transition are \emph{terminal}.
\item[{II}.] \label{II}
All states in $S$ are
(first-order) structures over the same finite vocabulary $\F$,
and $X$ and $\tau(X)$ share the same domain for any $X\in S$.
For
convenience, we treat relations as truth-valued functions and
refer to structures as algebras,
and let $t_{X}$ denote the value of term $t$ as interpreted in
state $X$.%
\footnote{All ``terms'' in this paper are ground (i.e.\ variable-free).}
The sets of states, initial states, and terminal states are each closed under isomorphism.
Moreover,
transitions respect isomorphisms. Specifically,
if $X$ and $Y$ are isomorphic, then either both are terminal or
else $\tau(X)$ and $\tau(Y)$ are also isomorphic
via the same isomorphism. 
\item[{III}.] \label{III}
There exists a fixed finite
set $T$ of \emph{critical} terms
 over $\F$ that fully determines the behavior of the algorithm.
Viewing any state $X$ over $\F$ with domain $D$
as a set of location-value pairs $f(a_{1},\ldots,a_{n})\mapsto a_{0}$,
where $f\in\F$ and $a_{0},a_{1},\ldots,a_{n}\in D$,
this means that whenever states $X$ and $Y$ \emph{agree} on $T$,
in the sense that $t_{X}=t_{Y}$ for every critical term $t\in T$,
either both are terminal states or else
 $\tau(X)\setminus X=\tau(Y)\setminus Y$.
\end{enumerate}
\end{defn}

For detailed support for this characterization of algorithms, see~\cite{ASM-Theorem-Gurevich,BSL}.
Clearly, we are only interested here in deterministic algorithms.
We use the adjective ``classical'' to clarify that, in the current study, we are leaving aside
new-fangled forms of algorithm, such as probabilistic,
parallel or interactive algorithms.


A classical algorithm may be thought of as a class of \emph{implementations}, each computing
some (partial) function over its state space. An implementation is determined by the choice
of representation for the values over which the algorithm operates, which is reflected in a choice of domain.

\begin{defn}[Implementation]\label{def:imp}
An \emph{implementation} is 
an algorithm  $\langle \tau,S,S_{0}\rangle$ restricted to a specific domain
$D$. Its states are those states $S\upharpoonright D$
with domain $D$; its \emph{input states} $S_{D}\subseteq S_{0}$ are those initial
states whose domain is $D$; its transition function $\tau$ is likewise
restricted.
 \end{defn}
\noindent
So  we may view implementations as computing a function
over its domain. 

In the following, we will always assume  
a predefined subset $I\cup\{z\}$ of the critical terms, called \emph{inputs} and \emph{output}, respectively. 
Input states may differ only on input values  and input values must cover  the whole domain.
Then we may speak of an algorithm $A$ with terminating run $X_0 \leadsto_A \cdots \leadsto_A X_N$
as computing $A ( y^{1}_{X_0},\ldots,y^{k}_{X_0}) = z_{X_N}$.
The presumption that an implementation accepts any value from its domain as
a valid input is not a limitation, because the outcome of an
implementation on undesired inputs is of no consequence.


The postulates in Definition~\ref{def:alg} limit transitions to be effective, in the sense of being programmable, as we just saw, but they place no constraints
on the contents of initial states. In particular, initial states may contain infinite, uncomputable
data.
To preclude that, we will need an additional assumption.

\begin{defn}[Basic]\label{def:basic}
We call an algebra $X$ over vocabulary $\F$ and with domain
$D$ \emph{basic} if $\F=K\uplus J$,
 $D$ is isomorphic to the Herbrand universe (free term algebra) over $K$, the \emph{constructors} of $X$,
 and  $t_{X}=s_{X}\neq  \Undef$ for at most  a finite number of terms $t$ and
$s$ over $K\uplus J$,
for some pervasive constant value \Undef.
An implementation is \emph{basic} if
all its initial states are basic
with respect to the same constructors.
\end{defn}

Constructors are the usual way of thinking of the domain values of computational models.
For example, strings over an alphabet $\{\mbox{a,b,\dots}\}$ are constructed from
a nullary constructor $\varepsilon$ and unary constructors a$(\cdot)$, b$(\cdot)$, etc.
The positive integers in binary notation may be constructed out of 
the nullary $\varepsilon$ and unary 0 and 1,
with the constructed string understood as the binary number obtained by prepending the digit 1.

\begin{defn}[Effectiveness \cite{CT_ASM}]\label{def:eff}
Let $X$ be an algebra over vocabulary $\F$ and domain
$D$. We call $X$ \emph{effective} over $\F=K\uplus C$
if $K$ constructs $D$ and
each of the operations in $C$ can be computed by an effective implementation over $K$.
In other words, $C$ is a set of effective oracles, obtained by bootstrapping from basic implementations.
An \emph{effective implementation} is a classical algorithm restricted to
 initial states that are all effective.
over the {same} partitioned vocabulary $\F=K\uplus C$.
\end{defn}

Clearly, the properties of being basic or effective are closed under the transition of algorithm (this follows from Postulate III).
Hence, any reachable state of basic (effective) implementation is also basic (effective, respectively). 

\section{Complexity}\label{sec:com}

Complexity of an algorithm  is classically measured as a number of single
steps required by execution, relative to the size of the initial data.
This requires an interpretation of the notions {}``initial data size''
and ``single step''. By a ``step'', we usually mean
a single step of some well-defined theoretical computational model,
like a Turing machine or RAM, implementing an algorithm over a chosen
representation of the domain. 

An effective implementation may simulate an effective algorithm over
a chosen representation of domain, but it still cannot count for a
faithful measure of a single step, since   its states are
allowed to contain infinite non-trivial information as an oracle;
unlike a  basic implementation.

Basic implementations provide an underlying model for effective ones (and thus are a faithful measure of a single step):
\begin{prop}\label{pro:bootstrap-assignmnet-eliminating}Let $P=\langle \tau,S,S_{0}\rangle $
be an effective implementation over $K\uplus C$. Then there
exists a basic implementation simulating $P$ over
some vocabulary $K\uplus J$. \end{prop}
\noindent
The proof uses the notion of \emph{simulation} defined in \cite{Simulation} and standard programming techniques of  internalizing operations by bootstrapping.

For example, if an effective implementation includes decimal multiplication among its bootstrapped operations,
then we do not want to count multiplication as a single operation (which would give  a ``pseudo-complexity'' measure),
but, rather, the number of basic decimal-digit operations, as would be counted in the basic simulation of the effective
implementation.

With a notion of single step in hand, we are only left to define a suitable notion of  input size.
Let $P=\langle\tau,S,S_{0}\rangle$ be an effective implementation with constructors $K$.
Recall from Definition~\ref{def:basic} that the domain
of each $X\in S_0$ is identified with the Herbrand universe over $K$. Thus,
domain elements may be represented as terms over the constructors $K$.
Now, we need to measure the size of input values $y$, represented as constructor terms. 
The standard way to do this would be to count the number of symbols $|y|$ in the constructor term for $y$.
The more conservative way is
to count the minimal number of constructors required to access
it, which we propose to do.
 For example, we want the size of $f(c,c)$ to be 2, not 3.

\begin{defn}[Size]\label{def:size} The \emph{(compact) size} of a
term $t$ over vocabulary $K$ is 
$\|t\|:=|\{s:s\mbox{ is a subterm of }t\}$.
 \end{defn} 

Still another issue to consider is this:  a domain may be constructible by infinitely many different finite sets of constructors, which affects the measurement of  size. We are accustomed to say that the size of $n\in\mathbb{N}$
is $\lg n$, relying on the binary representation of natural numbers.
This, despite the fact that the implementation itself may use 
tally (unary) notation or any other representation. 
Consider now that somebody states that she has an effective implementation
over $\mathbb{N}$,  working under the supposition
that the size of $n$ ought to be measured by $\log\log n$. Should this be legal? We 
neither allow nor reject such statements with blind eyes, but 
require  justification. 

Switching representations of the
domain, one actually changes the vocabulary and thus the whole description
of the implementation.  Still, we want to recognize the result as
being the ``same'' implementation, doing the same job, even over the different
vocabularies. 

\begin{defn}[Valid Size] Let $A$ be an effective
implementation over domain $D$. A function $f:D\rightarrow\mathbf{N}$
is a \textit{valid size} for elements of $D$ if there is an effective
implementation $B$ over  $D$ such that
$A$ and $B$ are computationally equivalent (each simulating the other) via some bijection $\rho$, 
such that $f(x)=|\rho(x)|$ for
all $x\in D$. \end{defn}

\section{Abstract State Machines}\label{sec:asm}\label{def:asm}

Abstract state machines (ASMs)~\cite{Lipari,ASM-Theorem-Gurevich,Generic} provide a perfect language for  descriptions
of algorithmic transition functions.
They consist of generalized assignment statements
$f(s^{1},\ldots,s^{k}):=u$,  conditional tests
$\If C \Then P$ or $\If C \Then P \Else Q$,
where $C$ is a Boolean combination of equations between terms,
and parallel composition.
A program as such defines a single transition;
it
is executed repeatedly, as a unit, until no assignments have their conditions are enabled.
If no assignments are enabled, then there is no next state.

A triplet $\langle\M,S,S_{0}\rangle$ is called  \textit{abstract
state machine (ASM)} if  $S_0$ are initial states and $S$ are states of an ASM program  $\M$, such that $\langle\M,S,S_{0}\rangle$ satisfy the conditions for being an algorithm given in  Definition \ref{def:alg}.
\begin{ignore}
 If $\M$  is an ASM-program defined over 
Denote by $\textrm{Alg}_{\F}$ the class of all algebras over $\F$. An ASM-program
$\M$ defines a partial transition function $\M:\textrm{Alg}_{\F}\rightharpoonup \textrm{Alg}_{\F}$
such that $\M(X)$ is the result of applying $\M$ to
$X$. 
Let $\M$ be an ASM program over $\F$ and 
let $S_{0}\subseteq S\subseteq \textrm{Alg}_{\F}$ be such that
$\langle\M,S,S_{0}\rangle$ is an algorithm in the sense of Definition~\ref{def:alg}.
Then $\langle\M,S,S_{0}\rangle$ will be called an \textit{abstract
state machine (ASM)}. 
\end{ignore}
Every algorithm is emulated step-by step, state-by-state by an ASM.

\begin{thm}[\cite{ASM-Theorem-Gurevich}]\label{thm:ESMisEffective-1}
Let $\langle\tau,S,S_{0}\rangle$ be an  algorithm over vocabulary
$\F$. Then there exists an ASM $\langle\M,S,S_{0}\rangle$ over the
same vocabulary, such that $\tau=\M\upharpoonright_{S}$,
with the terms (and subterms) appearing in the ASM program serving as critical terms.
\end{thm} 

\begin{defn}[ESM] 
An \emph{effective state machine (ESM)}
is an effective implementation of an ASM $\M$.
\end{defn} 

Constructors are part and parcel of the states, though they   need not appear in an ESM program.

\section{Simulation}\label{sec:sim}


We know from~\cite[Theorem 3]{CT_ASM}
that for any effective model there is a string-representation of its domain
under which each effective implementation has a Turing machine that 
simulates it, and---by the same token---there are
RAM simulations.
Our goal is to prove that it can be done at polynomial cost.
We  describe a RAM algorithm satisfying these conditions. The result
will then follow from the standard poly-time (cubic) connection between TMs and RAMs. 
First, we need to choose an appropriate RAM representation
for our domain of terms.

For  term $t$, we denote  the minimal graph representing it by  $\Gr{t}$ 
and  the quantity of RAM memory required to store it by   $|\Gr{t}|$. 
These memory cells will each contain a small constant, indicating 
a vertex label or a pointer, corresponding to an edge in the graph.
Note that since  $\Gr{t}$  is minimal, it does not contain repeated factors.
To prevent repeated factors  not just in one term, but in the whole state, 
we merge the individual term graphs into one big graph and call 
the resulting  ``jungle''  a  \emph{tangle}  (see~\cite{TermGraph}). 
The tangle will maintain the constructor-term values of all the critical terms of the algorithm. 
Consider, for example, the 
natural way to merge terms $t=f(c,c)$ and $s=g(c,c)$, where $c$
is a constant. The resulting dag $G$ has three vertices, labeled $f,g,c$.
Two edges  point from $f$ to $c$ and the other two  
 from $g$ to $c$.
Our two terms
may be represented as pointers to the appropriate vertex: $G(t)$
refers to the $f$ vertex and $G(s)$ to $g$, where
we are using the notation $G(t)$ to refer to the vertex in $G$ that represents the term $t$.

\begin{prop} 
For any tangle $G$ of terms over a finite vocabulary, we have
$|E(G)|=O(|V(G)|)$.
\end{prop}

Let  $\langle\M,S,S_{0}\rangle$ be a basic ESM over vocabulary $\F=K\uplus J$, with 
 input terms $I\subseteq J$, and critical terms $T=\{t^1,\ldots,t^m\}$, including all their subterms,
 ordered from \emph{small to big}.
Also, let $X_{0}\leadsto_\M X_{1}\leadsto_\M\cdots$ be
some run of $\M$, for which we let  $\Gr{T_i}$ denote  the
tangle of the domain values $\{t_{X_{i}}: t\in T\}$ of the critical terms in the $i$-th state $X_i$.
For $\bar t$, a finite sequence or set of terms,  we  use the abbreviation $\|\bar t\|=\sum_{s\in\bar t}\|s\|$.

One transition of ESM involves a bounded number of comparisons of the  values of critical terms. The cost for each is constant:

\begin{prop} \label{pro:Gt is linear size}\label{pro:ComparingComplexity}
Let $\Gr T$ be a critical tangle and let $s$ and $t$  be critical terms in $T$. 
Therefore, the question whether $\Gr{t}=\Gr{s}$,  
is decidable in constant number of RAM-operations of logarithmic word size.
\end{prop} 

One transition of an ESM involves a bounded number of assignments. The cost
of each assignment is linear:

\begin{prop} \label{pro:AssignmComplexity}
Let $t=f(\bar t)$ be a term over vocabulary $K$.
Then $\Gr{t}$
can be constructed using  $O(\|\bar t\|)$ RAM-operations  of logarithmic word size.
\end{prop} 


Combining the previous propositions together, we may conclude:

\begin{prop}\label{prop: size-growth} 
The critical tangles grow by a constant amount in each step.
So,  $|\Gr{T_i}|= O(|\Gr{T_0}|+i).$
\end{prop}


\begin{thm}[ Initial States]\label{Simulating-initial-state} 
Given term-graphs $\Gr{\In}$ for the inputs $\In$ in an initial state
 $X_{0}$, Algorithm~\ref{algo2} constructs the initial critical tangle $\Gr{T_0}$ in $O(\|\In\|)$ steps.
\end{thm} 

\begin{algorithm}[H]
\caption{}\label{algo2}

\begin{itemize}
\item $\mbox{for }i=1,\ldots,m$ 

\begin{itemize}
\item let $t^{i}=f(s^{1},\ldots,s^{\ell})$
\item if all $s^{j}$ are defined, then

\begin{itemize}
\item if $f\in K$,  create $f(s^{1},\ldots,s^{\ell})$, as described
in Proposition \ref{pro:AssignmComplexity}
\item  if $f\notin K$, then

\begin{itemize}
\item if found $r^{1},\ldots, r^{\ell},r\in \Gr{T}$ such that $\Gr{r^{j}}=\Gr{s^{j}}$
for all $j$ and $r=f(r^{1},\ldots,r^{\ell})$ is defined, then copy the content of $r$ to $t^{i}$
\end{itemize}
\end{itemize}
\end{itemize}
\end{itemize}
\end{algorithm}

\begin{thm}[Transitions]\label{Simulating-the-transition} 
Algorithm~\ref{alg3} computes the successor tangle
$\Gr{T_{i+1}}$ from $\Gr{T_{i}}$ in time linear in $|\Gr{T_{i}}|$.
\end{thm} 

\begin{algorithm}[H]
\caption{}\label{alg3}

\begin{itemize}
\item for each critical term $t\in T$ create a new pointer $t'$ to point to its new
value
\item for each possible assignment in the ESM, do the following:

\begin{itemize}
\item if all guards evaluate to \True{}, then
\item for the enabled assignment $f(s^{1},\ldots,s^{\ell}):=s$ do

\begin{itemize}
\item if $s$ is {}\Undef{} then $f'(s^{1},\ldots,s^{\ell})$
is also \Undef{}
\item if some $s^{i}$ is {}\Undef{} then $f'(s^{1},\ldots,s^{\ell})$
is also \Undef{}
\item otherwise, if $f\in K$ then 

\begin{itemize}
\item set $f'(s^{1},\ldots,s^{\ell})$ to point to the
graph constructed as described in Proposition \ref{pro:AssignmComplexity}
\end{itemize}
\item whereas, if $f\notin K$, then if found $r^{1},\ldots r^{\ell},r\in \Gr{T}$ such that
$\Gr{r^{j}}=\Gr{s^{j}}$ for all $j$ and $r=f(r^{1},\ldots,r^{\ell})$
is defined, then

\begin{itemize}
\item if $f'(s^{1},\ldots,s^{\ell})$ is not {}\Undef, 
set $f'(s^{1},\ldots,s^{\ell})$ to point to a copy
of $\Gr{r}$
\end{itemize}
\end{itemize}
\end{itemize}
\end{itemize}
\end{algorithm}

The result we set out to achieve now follows.

\begin{thm}[Simulating ESMs]\label{Simulating-ESM}
Any effective implementation with
complexity $T(n)$, with respect to a valid size measure,
can be simulated by a RAM in order
$n+n T(n)+T(n)^{2}$ steps,
with a word size that grows to order $\log T(n)$.\end{thm} 

\section{Summary}\label{sec:sum}

We have shown---as has been conjectured---that every effective implementation,
regardless of what data structures it uses, can be simulated by a
Turing machine, with at most polynomial overhead in time complexity.
Specifically, we have shown that any algorithm running on an effective sequential model can be simulated, independent of the problem, 
by a single-tape Turing machine with a
quintic overhead: quadratic for the RAM simulation and another cubic for a TM simulation of the RAM \cite{RAM}.

To summarize the argument in a nutshell:
Any effective algorithm is behaviorally identical to an abstract state machine operating over a domain that is isomorphic to some Herbrand universe,
and whose term interpretation provides a valid measure of input size.
That machine is also behaviorally identical to one whose domain consists of maximally compact dags, labeled by constructors.
Each basic step of such a machine, counting also the individual steps of any subroutines,
increases the size of a fixed number of such compact dags by no more than a constant number of edges.
Lastly, each machine step
can be simulated by a RAM that manipulates those dags in time that is linear in the size of the stored dags.

It remains to be seen whether it may be possible to improve the complexity of the simulation.

\bibliography{models}
\bibliographystyle{eptcs}

\end{document}